\documentclass{llncs}
\usepackage{llncsdoc}
\usepackage{graphicx}
\usepackage{afterpage}
\usepackage[caption=false]{subfig}
\usepackage{multirow}
\usepackage{array}
\usepackage{xcolor}
\usepackage{float}
\usepackage[normalem]{ulem}
\usepackage{listings}
\let\oldFootnote\footnote
\newcommand\nextToken\relax

\renewcommand\footnote[1]{%
    \oldFootnote{#1}\futurelet\nextToken\isFootnote}

\newcommand\isFootnote{%
    \ifx\footnote\nextToken\textsuperscript{,}\fi}

\newcolumntype{C}[1]{>{\centering\let\newline\\\arraybackslash\hspace{0pt}}m{#1}}
\newcolumntype{A}{>{\centering}m{0.588cm}}

\begin{document}
\title{FINJ: A Fault Injection Tool for HPC Systems}

\author{}
\institute{}

 \author{
 Alessio Netti\inst{1}, Zeynep Kiziltan\inst{1}, Ozalp Babaoglu\inst{1} \\ Alina S\^irbu\inst{2}, Andrea Bartolini\inst{3}, Andrea Borghesi\inst{3}}
        
 \institute{Department of Computer Science and Engineering\\ University of Bologna, Italy \\ 
     \email{\{alessio.netti, zeynep.kiziltan, ozalp.babaoglu\}@unibo.it}
     \and
     Department of Computer Science\\ University of Pisa, Italy \\ 
     \email{alina.sirbu@unipi.it}
     \and
     Department of Electrical, Electronic and Information Engineering\\ 			University of Bologna, Italy \\
     \email{\{andrea.bartolini, andrea.borghesi\}@unibo.it}
}

\maketitle

\begin{abstract}
We present FINJ, a high-level fault injection tool for High-Performance Computing (HPC) systems, with a focus on the management of complex experiments. FINJ provides support for custom workloads and allows generation of anomalous conditions through the use of fault-triggering executable programs. FINJ can also be integrated seamlessly with most other lower-level fault injection tools, allowing users to create and monitor a variety of highly-complex and diverse fault conditions in HPC systems that would be difficult to recreate in practice. FINJ is suitable for experiments involving many, potentially interacting nodes, making it a very versatile design and evaluation tool.
\end{abstract}


\section{Introduction}
\label{section:introduction}

\paragraph{Motivation.} High-Performance Computing (HPC) systems have become indispensable for economic growth and scientific progress in our modern society. As the performance of HPC systems increases, the value of the results they produce increases through higher-fidelity simulations, better predictive models and analysis of greater quantities of data. The resulting techniques, policy decisions and vastly-improved manufacturing processes in areas such as agriculture, engineering, transportation, materials, energy, health care, security and the environment are bound to impact most aspects of our lives.  Today, HPC systems are also being used as fundamental ``instruments" to achieve groundbreaking results in basic sciences ranging from particle physics to cosmology.
Yet, many important problems in various fields remain unsolvable with current computational resources. \emph{Exascale} HPC systems, capable of \(10^{18}\) operations per second, are believed to be essential for solving such problems~\cite{ashby2010opportunities}. Reaching exascale performance is the \emph{moonshot} for modern HPC systems with many nations and companies engaged in an \emph{arms race} towards achieving it.

Exascale systems, when they arrive, will come at a significant cost: scaling current technologies to exascale performance through massive parallelism will result in systems that have prohibitively-high levels of power consumption~\cite{villa2014scaling} and excessively-high failure rates~\cite{cappello2014toward}. Thus, to be usable in production environments with acceptable \emph{Quality of Service} levels, exascale systems need to improve their power efficiency and resiliency by several orders of magnitude.

In our terminology, a \emph{fault} is defined as an anomalous behavior at the software or hardware level that can lead to illegal system states (\emph{errors}) and, in the worst case, to service interruptions (\emph{failures})~\cite{gainaru2015errors}.  In this paper, we limit our attention to improving the resiliency of HPC systems through  the use of mechanisms for predicting, detecting and preventing errors and failures. An important technique in this endeavor is \emph{fault injection}: the deliberate triggering of faults in a system so as to observe their behavior in a controlled environment, enable development of new prediction and response techniques and testing of existing ones~\cite{hsueh1997fault}. For fault injection to be effective, dedicated tools are necessary, allowing users to trigger complex and realistic fault scenarios in a reproducible manner. 

\paragraph{Related Work.}
\label{section:relatedwork}
Fault injection for prediction and detection purposes has been a topic of great interest in recent years. In~\cite{tuncer2017diagnosing,guan2012cda,guan2013adaptive,ferreira2008characterizing}, the authors employed software-based fault injection techniques to observe the behavior and performance variations of HPC systems in anomalous conditions, and to detect such faults using system performance metrics. However, while characterizing the fault-simulating programs that were used, these works do not focus on the tools used to inject and coordinate the faults themselves in the system. 

Several studies have proposed fault injection tools with varying levels of abstraction. Calhoun et al.~\cite{calhoun2014flipit} devised a compiler-level fault injection tool focused on memory bit-flip errors, targeting HPC applications. De Bardeleben et al.~\cite{debardeleben2011experimental} proposed a logic error-oriented fault injection tool. This tool is designed to inject faults in virtual machines, by exploiting emulated machine instructions through the open-source virtual machine and processor emulator (QEMU). Both works focus on low-level fault-specific tools and do not provide functionality for the injection of complex workloads, and for the collection of produced data, if any.

Stott et al.~\cite{stott2000nftape} proposed NFTAPE, a high-level and generic tool for fault injection. This tool is designed to be integrated with other fault injection tools and triggers at various levels, allowing for the automation of long and complex experiments. The tool however has aged considerably, and is not publicly available. A similar fault injection tool was proposed by Naughton et al.~\cite{naughton2009fault}, however, to the best of our knowledge, it has never progressed past the prototype stage and is also not publicly available. Moreover, both tools require users to write a fair amount of wrapper and configuration code, resulting in a complex setup process. The Gremlins Python package\footnote{\url{https://github.com/toddlipcon/gremlins}} also supplies a high-level fault injector. However, it does not support workload or data collection functionalities, and experiments on multiple nodes cannot be performed.

Joshi et al.~\cite{joshi2011prefail} introduced the PREFAIL tool, which allows for the injection of failures at any code entry point in the underlying operating system. This tool, like NFTAPE, employs a coordinator process for the execution of complex experiments. It is targeted at a specific type of fault (code-level errors) and does not permit performing experiments focused on performance degradation and interference, among other fault types. Similarly, the tool proposed by Gunawi et al.~\cite{gunawi2011fate}, named FATE, allows the execution of long experiments; furthermore, it is focused on reproducing specific fault sequences, simulating real scenarios. Like PREFAIL, it is limited to a specific fault type, namely I/O errors, thus greatly limiting its scope.

\paragraph{Contributions.} The main contribution of this paper is the design and implementation of FINJ, an easy-to-use open-source Python tool for fault injection targeted at HPC systems, with workload management capabilities. A relevant feature of FINJ is the possibility of seamless integration with other injection tools targeted at specific fault types, thus enabling users to coordinate faults from different sources and different system levels. By using FINJ's \emph{workload} feature, users can also specify lists of applications to be executed and faults to be triggered on multiple nodes at specific times with specific durations. FINJ thus represents a high-level, flexible tool, enabling users to perform complex and reproducible experiments, aimed at revealing the complex relations that may exist between faults, application behavior and the system itself. FINJ is also extremely easy to use: it can be set up and executed in a matter of minutes, and does not require the writing of additional code in most of its usage scenarios.
To the best of our knowledge, FINJ is the first portable, open-source tool that allows users to perform and control complex injection experiments, that can be integrated with heterogeneous fault types and that includes workload support, while retaining ease of use and a quick setup time.

\paragraph{Organization.} The rest of the paper is structured as follows. In Section~\ref{section:architecture}, we describe the FINJ architecture (Section~\ref{section:working}), its components (Section~\ref{section:component}) and their implementation (Section~\ref{section:implementation}). In Section~\ref{section:casestudy}, we present a simple use case to show how FINJ can be deployed, while Section~\ref{section:conclusions} concludes the paper. 

\section{FINJ Architecture}
\label{section:architecture}

In this Section we discuss how fault injection is achieved in FINJ. We then present its architecture, together with some implementation details. Due to its portable and modular nature, customizing FINJ for different purposes is easy.

\subsection{Architecture Overview}
\label{section:working}

Fault injection in FINJ is achieved through \emph{tasks} that are executed on target nodes: each task corresponds to a particular application, which can either be a benchmark program or a fault-triggering program. As demonstrated in \cite{stott2000nftape}, this approach allows for the integration in FINJ of any low-level fault injection framework that can be triggered by using an executable program or a shell script. A task is defined by the following attributes:

\begin{itemize}
\item \emph{args}: the full shell command required to run the selected task. The command must refer to an executable file that can be accessed from the target hosts;
\item \emph{timestamp}: the time in seconds at which the task must be started, relative to the starting time of the injection session;
\item \emph{duration}: the task's \emph{maximum allowed} duration, expressed in seconds, after which it will be abruptly terminated. This duration can serve as an \emph{exact} duration as well, with FINJ restarting the task if it finishes earlier, and terminating it if it lasts more. This behavior depends on the FINJ configuration (see Section~\ref{section:component}). A duration of 0 implies that the task is always allowed to run until its termination;
\item \emph{isFault}: defines whether the task corresponds to a fault-triggering program, or to a benchmark application;
\item \emph{seqNum}: a sequence number used to uniquely identify the task;
\item \emph{cores}: the list of CPU cores that the task is allowed to use on target nodes, enforced through a \emph{NUMA Control} policy \cite{lameter2013numa}; this attribute is optional. 
\end{itemize}

A set of tasks defines a \emph{workload}, which is a succession of scheduled fault and benchmark executions at specific times, reproducing a realistic working environment for the fault injection process. A particular execution of a given workload then constitutes an \emph{injection session}. Many fault programs are supplied with FINJ, allowing users to experiment with a variety of anomalies out-of-the-box.

\begin{figure}[t]
  \centering
	\includegraphics[width=0.9\textwidth]{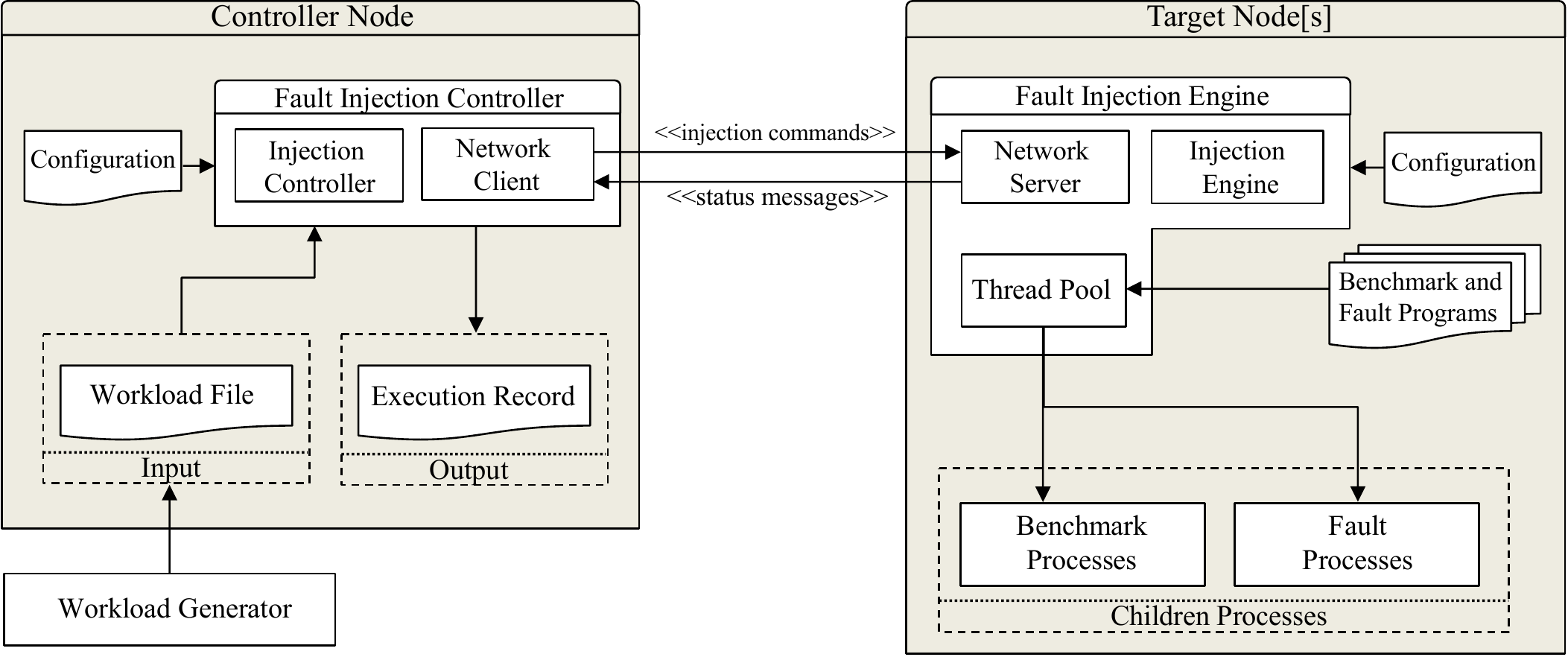}
  \caption{Architecture of the FINJ tool showing the division between a controller node (left) and a target node (right).}\label{fig:architecture}
\end{figure} 

FINJ consists of two basic components: a fault injection \emph{controller}, and a fault injection \emph{engine}. The two components correspond to processes that must be run on the nodes subject to injection experiments. Communication between them is achieved through TCP \emph{sockets} using a simple message-based protocol. The high-level structure of the FINJ architecture is illustrated in Figure~\ref{fig:architecture}.

\paragraph{FINJ Controller.} The controller is the orchestrator of the injection process, and should be run on an external node that is not affected by the faults. The controller maintains connections to all nodes involved in the injection session, which run fault injection \emph{engine} instances and whose addresses are specified by users when launching the program. Therefore, injection sessions can be performed on multiple nodes at the same time. The controller reads task entries from the selected workload: the reading process is incremental, and tasks are read in real-time a few minutes before their expected execution, according to their relative time-stamp. For each task the controller sends a command to all target hosts, instructing them to start the new task at the specified time. Finally, the controller collects all status messages produced by the target hosts, and stores them in a separate file for each host. These status messages are related to the start and termination of each single task, besides status changes in the host (for example, when connection is lost and re-established).

\paragraph{FINJ Engine.} The engine is structured as a daemon, perpetually running on nodes that are expected to be subject to injection sessions. The engine waits for task commands to be received from remote controller instances. Engines can be connected to multiple controllers at the same time, and status messages will be sent to all of them. However, task commands are accepted from one controller at a time, which is defined as the \emph{master} of the injection session. The engine manages received task commands by assigning them to a dedicated thread from a \emph{pool}. The thread manages all aspects related to the execution of the task, such as spawning the necessary subprocesses and sending status messages to controllers when relevant events (such as the start or termination of the task) occur. Whenever a fault causes a target node to crash and reboot, controllers are able to re-establish and recover the previously running injection session, given that the engine is set up to be executed at boot time on the target node.

\subsection{Components}
\label{section:component}

FINJ is based on a highly modular architecture, and therefore it is very easy to customize single components in order to add or tune features.

\paragraph{Network.} Engine and controller instances communicate through a network layer in the FINJ tool. Communication is achieved through a simple message-based protocol employing TCP sockets. This design choice is motivated by the fact that the volume of data sent during injection sessions is extremely low, while high reliability is a desirable quality. Users can still integrate their preferred transport method with little effort, thanks to FINJ's highly modular nature.

Specifically, a message \emph{client} and \emph{server} were implemented: clients are used by FINJ controllers in order to connect to servers hosted on FINJ engine instances. Messages can then be either \emph{commands}, related to single tasks and imposed by controllers, or \emph{status} messages, which are sent by engines and are related to status changes in their system. All messages are in the form of \emph{dictionaries}. This component also handles resiliency features such as automatic re-connection from clients to servers, since temporary connection losses are to be expected in a fault injection context.

\paragraph{Thread Pool.} Task commands in FINJ engines are assigned to a thread in a pool as they are received: each thread manages all aspects of a task assigned to it. Specifically, the thread sleeps until the scheduled starting time of the task (according to its time-stamp); then, it spawns a subprocess running the specified task, and sends a message to all connected controllers to inform them of the event. At this point, the thread waits for the task's termination, depending on its duration and on the current configuration. Finally, the thread sends a new status message to all connected hosts informing them of the task's termination, and returns to sleep. The amount of threads in the pool, which is a configurable parameter, determines the maximum number of tasks that can be executed concurrently. Since threads in the pool are started only once during the engine's initialization, and wake up for minimal amounts of time when a task needs to be started or terminated, we expect their impact on performance to be negligible.

\paragraph{Input and Output.} In FINJ, input and output of all data related to injection sessions are performed by controller instances, and are handled by \emph{reader} and \emph{writer} entities. By default, these employ the CSV format, which was chosen due to its extreme simplicity and generality, but they can be easily customized by users for other formats. \emph{Input} in FINJ is constituted by \emph{workload} files: as mentioned in Section~\ref{section:working}, these files include one entry for each task that must be executed in the injection session. Using the CSV format makes workload files extremely readable, and manually writing workloads corresponding to highly specific test cases can be easily achieved as well. FINJ \emph{output}, instead, is made up of two parts. The first is the \emph{execution log}, which contains entries corresponding to status changes in the target node, namely the start and termination of tasks, errors that are encountered if any, and connection loss or recovery events. The second part of FINJ output is related to tasks: all output text written to the \emph{stdout} or \emph{stderr} channels during their execution, if any, is reported to controllers, and is stored in separate plain-text files in a directory alongside the main output file, each named according to the task's name and sequence number.

\paragraph{Configuration.} The FINJ tool's runtime behavior is customizable by means of a configuration file. This file is in JSON format and includes several options that alter the behavior of either controller or engine instances. Among the basic options, it is possible to specify the listening TCP port for engine instances, and the list of addresses of target hosts, to which controller instances should connect at launch time. The latter is useful when injection sessions must be performed on large sets of nodes, whose addresses can be conveniently stored in a file. More complex options are also available: for instance, it is possible to define a series of commands corresponding to tasks that must be launched together with FINJ, and must be terminated with it. This option proves especially useful when users wish to set up monitoring frameworks, such as the \emph{Lightweight Distributed Metric Service} (LDMS) \cite{agelastos2014lightweight}, to be launched together with FINJ in order to collect system performance metrics during injection sessions. 

\paragraph{Workload Generation.} While writing workload files manually is possible, this is time-consuming and not desirable for long injection sessions. Therefore, we implemented in FINJ a \emph{workload generation} tool, which can be used to automatically generate workload files with certain statistical features, while trying to combine flexibility and ease of use. The workload generation process is controlled by three parameters: a maximum \emph{time span} for the total duration of the workload expressed in seconds, a statistical distribution for the \emph{duration} of tasks, and another one for their \emph{inter-arrival} times. These distributions are separated in two sets, for fault and benchmark tasks, thus amounting to a total of four. They can be either specified analytically by the user or can be fitted from real data, thus reproducing realistic behavior.

A workload is composed as a series of fault and benchmark tasks that are selected from a list of possible shell commands. To control the composition of workloads, users can optionally associate to each command a probability for its selection during the generation process, and a list of CPU cores for its execution, as explained in Section~\ref{section:working}. By default, commands are picked uniformly. Having defined its parameters, the workload generation process is then fairly simple: tasks are randomly generated in order to achieve statistical features close to those specified as input, and are written to an output CSV file, until the maximum imposed time span is reached. Alongside the full workload, a \emph{probe} file is also produced: this workload file contains one entry for each task type, all with a short fixed duration, and represents a lightweight workload version. This file can be used during the setup phase to test the correct configuration of the system, making sure that all tasks are correctly found and executed on the target hosts, without having to run the entire heavy workload. 

\subsection{Implementation}
\label{section:implementation}

FINJ is implemented in Python, an object-oriented, high-level interpreted programming language\footnote{\url{https://www.python.org/events/python-events/}}, and can be used on all major operating systems. All FINJ dependencies are included in the Python distribution, and the only optional external dependency is the \emph{scipy} package, which is needed for the workload generation functionality. The source code is publicly available on GitHub\footnote{\url{https://github.com/AlessioNetti/fault\_injector}} under the MIT license, together with its documentation, usage examples and several fault-triggering programs. FINJ works on Python versions 3.4 and above.

\begin{figure}[t]
  \centering
	\includegraphics[width=0.95\textwidth]{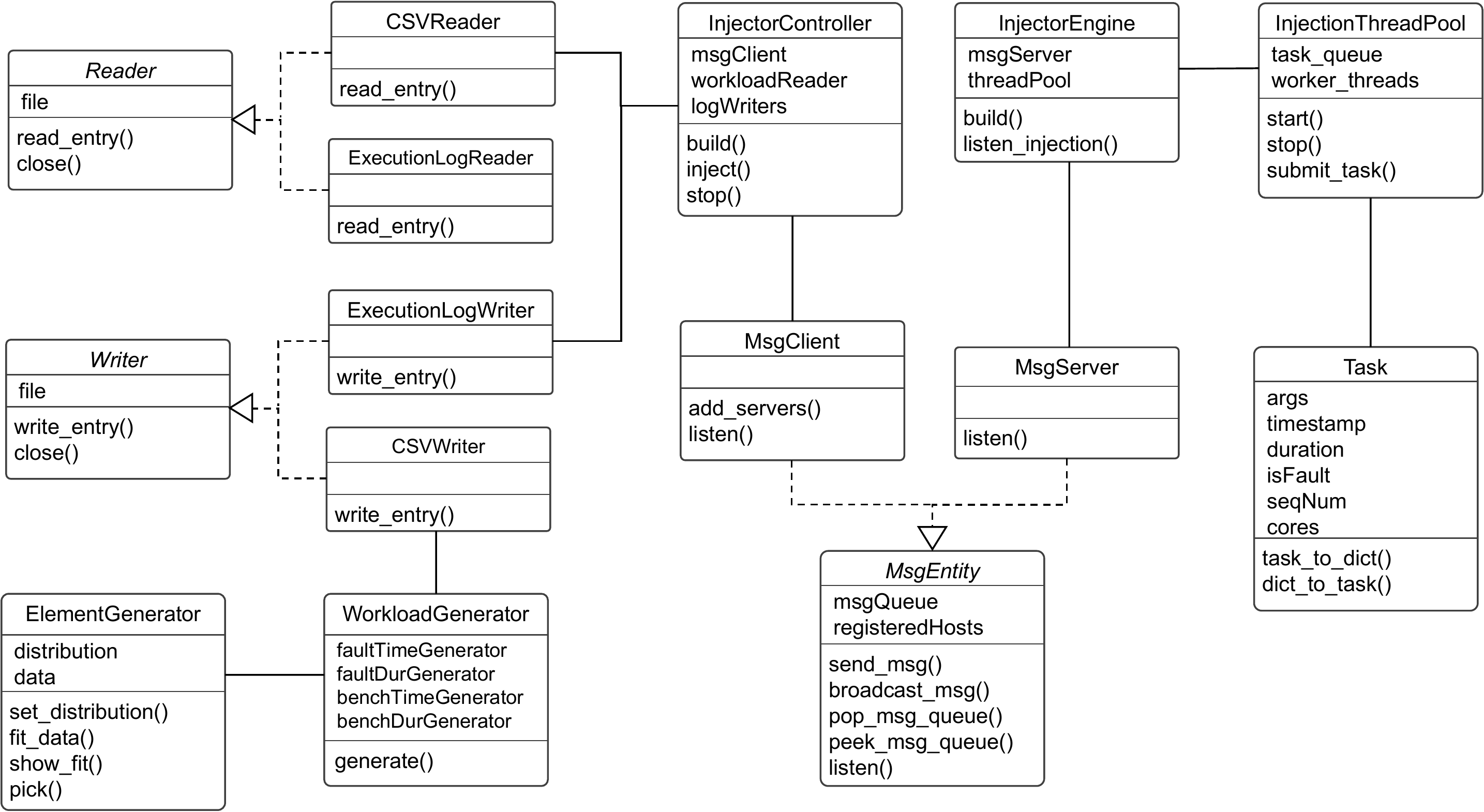}
  \caption{Class diagram of the FINJ tool.}\label{fig:classdiagram}
\end{figure} 

In Figure \ref{fig:classdiagram} we illustrate the class diagram for the FINJ tool. The \emph{engine} and \emph{controller} entities are respectively represented by the \emph{InjectorEngine} and \emph{InjectorController} classes. Users can instantiate these classes and start injection sessions directly, by using the \emph{listen} method to put the engine in listening mode, and the \emph{inject} method of the controller, which allows to start the injection session itself. However, scripts are supplied with FINJ to create controller and engine instances from a command-line interface, simplifying the process. This method will be discussed in Section~\ref{section:casestudy}. The \emph{InjectionThreadPool} class, instead, supplies the thread pool implementation used to execute and manage tasks.

The network layer of the tool is represented by the \emph{MsgClient} and \emph{MsgServer} classes, which implement the message and queue-based client and server used for communication. Both classes are implementations of the \emph{MsgEntity} abstract class, which provides the interface for sending and receiving messages, and implements the basic mechanisms that regulate the access to the underlying queue.

Input and output are instead handled by the \emph{Reader} and \emph{Writer} abstract classes and their implementations: \emph{CSVReader} and \emph{CSVWriter} handle the reading and writing of workload files, while \emph{ExecutionLogReader} and \emph{ExecutionLogWriter} handle execution logs generated by injection sessions. Since these classes are all implementations of abstract interfaces, it is easy for users to customize them for different formats. Tasks are modeled by the \emph{Task} class that contains all attributes specified in Section~\ref{section:working}.

Lastly, access to the workload generator is provided through the \emph{WorkloadGenerator} class, which is the interface used to set up and start the generation process. This class is backed by the \emph{ElementGenerator} class, which offers basic functionality for fitting data and generating random values. This class acts as a wrapper on scipy's \emph{rv\_continuous} class, which generates random variables.

\section{Using FINJ}
\label{section:casestudy}

In this Section we demonstrate the flow of execution of FINJ through a concrete example carried out on a real HPC node and provide insight on its overhead.

\subsection{Sample Execution}
\label{section:execution}

\begin{figure}[t]
\begin{lstlisting}[frame=tb]
timestamp;duration;seqNum;isFault;cores;args 
0;1723;1;False;0-7;./hpl lininput
355;244;2;True;6;sudo ./cpufreq 258
914;291;3;True;4;./leak 316
\end{lstlisting}
\caption{A sample CSV workload that can be used with FINJ.}
\label{fig:inputsample}
\end{figure}

In this Section we will consider a sample fault injection session carried out using FINJ. The employed workload file, named \emph{sample.csv}, is illustrated in Figure~\ref{fig:inputsample}. The test was carried out on one node of an HPC system equipped with two Intel Xeon E5-2630 v3 CPUs, each with 8 cores, 128GB of RAM, and running CentOS 7.3. The \emph{finj\_engine} and \emph{finj\_controller} Python scripts are supplied with FINJ to start \emph{engine} and \emph{controller} instances respectively. Their usage is explained on the GitHub repository for the tool, together with all configuration options.

In this workload, the first task is the Intel Distribution\footnote{\url{https://software.intel.com/en-us/mkl-windows-developer-guide-overview-of-the-intel-distribution-for-linpack-benchmark}} for the well-known \emph{High-Performance Linpack} (HPL) benchmark, optimized for Intel Xeon CPUs. This task starts at time 0 in the workload, and has a maximum allowed duration of 30 minutes. The following two tasks are fault-triggering programs: \emph{cpufreq} uses the Intel P-State driver in the Linux kernel\footnote{\url{https://www.kernel.org/doc/Documentation/cpu-freq/intel-pstate.txt}} to dynamically reduce the maximum allowed CPU frequency, emulating performance degradation, while \emph{leak} \cite{tuncer2017diagnosing} creates a memory leak in the system, eventually using all available RAM. The cpufreq program requires appropriate permissions, so that users can access the files controlling Linux CPU governors. The HPL benchmark was run with 8 threads, pinned on the first 8 cores of the machine, while the cpufreq and leak tasks were forced to run on cores 6 and 4 respectively. Also note that the tasks must be available at the specified path on the systems running the FINJ engine, which in this case is relative to the location of the launching script.

Having defined the workload, the injection engine and controller must be started. Using the default configuration, and supposing that the test must be performed locally, this can be accomplished with the two following commands:

\begin{lstlisting}[language=bash]
python finj_engine -p 30000 &
python finj_controller -w sample.csv -a localhost:30000
\end{lstlisting}

\begin{figure}[t]
\begin{lstlisting}[frame=tb]
timestamp;type;args;seqNum;duration;isFault;cores;error
1529172604;command_session_s;None;None;None;None;None;None
1529172624;status_start;./hpl lininput;1;1723;False;0-7;None
1529172979;status_start;sudo ./cpufreq 258;2;258;True;6;None
1529173237;status_end;sudo ./cpufreq 258;2;258;True;6;None
1529173538;status_start;./leak 316;3;316;True;4;None
1529173855;status_end;./leak 316;3;316;True;4;None
1529174347;status_end;./hpl lininput;1;1723;False;0-7;None
1529174348;command_session_e;None;None;None;None;None;None
\end{lstlisting}
\caption{A sample output file produced by FINJ after an injection session for the workload specified in Figure \ref{fig:inputsample}.}
\label{fig:outputsample}
\end{figure}

In the code above, the \emph{-p} argument indicates the listening TCP port for the engine instance. The \emph{-a} argument is instead the list of engine addresses to which the controller should connect, and \emph{-w} is the path of the CSV workload file to be injected.
The controller instance will then connect to the engine and start executing the workload, storing all output in a unique CSV file for each target host. When this process is finished, the controller terminates. The output CSV files for our example have the format shown in Figure \ref{fig:outputsample}: each entry represents a status change event, which in this case is the start or termination of tasks, and is flagged with its absolute time-stamp on the target host. In addition, we also find an \emph{error} field, detailing possible errors that were encountered. Note that the file is opened and closed by session \emph{start} and \emph{end} entries: the presence of these ensures that the injection process did not encounter errors and that the entire workload was processed successfully. It can be clearly seen from this experiment how easily a FINJ experiment can be configured and started on multiple cores.

\begin{figure}[t]
  \centering
	\includegraphics[width=0.95\textwidth]{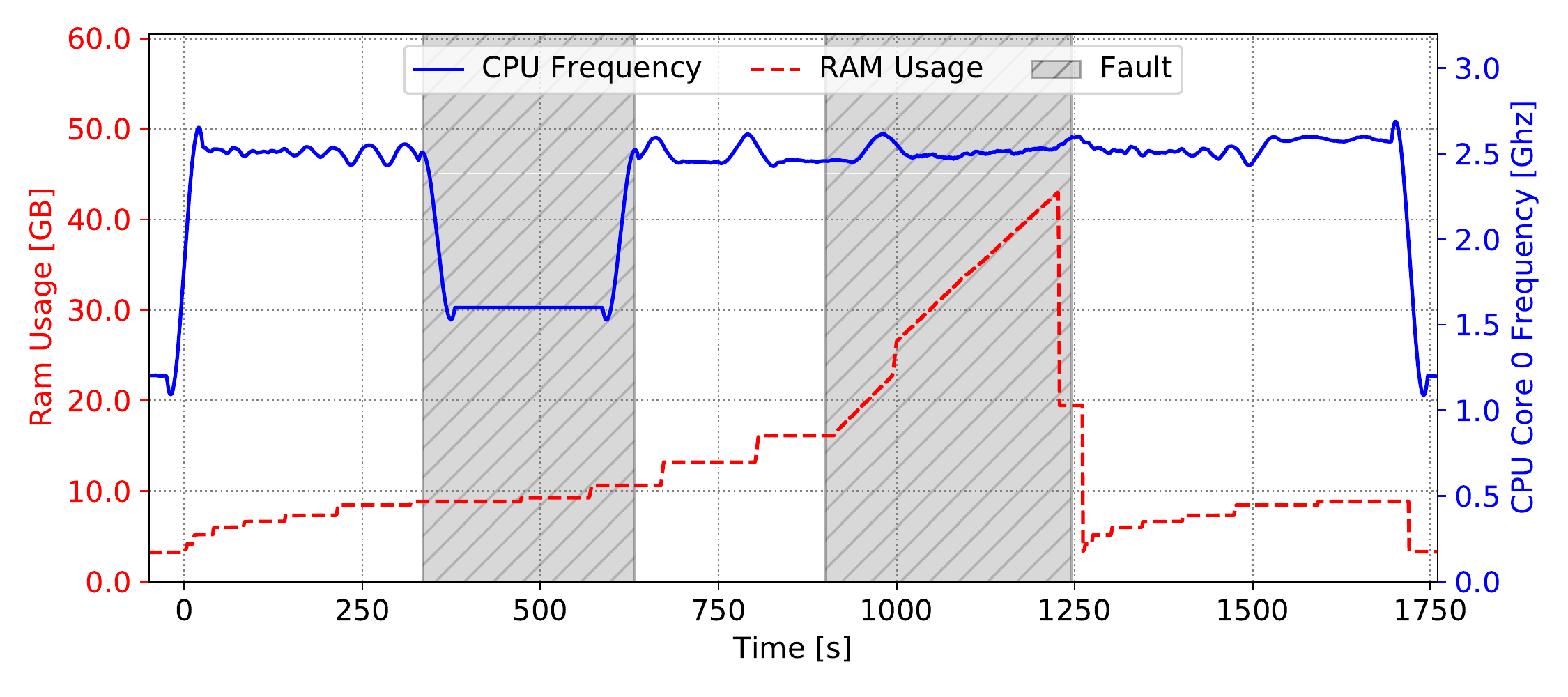}
  \caption{CPU Frequency and RAM Usage, as monitored on the target system during the sample injection session.}\label{fig:execution}
\end{figure}

At this point, the data generated by FINJ can be easily compared with other data, for example performance metrics collected through a monitoring framework, in order to perform fault detection studies or simply to better understand the system's behavior under faults. For this test, we used the LDMS framework \cite{agelastos2014lightweight} to collect resource usage and performance metrics on the target host at each second, for the duration of the injection session. In Figure \ref{fig:execution} we show the total RAM usage and the CPU frequency of core 0. The benchmark's profile is simple, showing a constant CPU frequency while RAM usage slowly increases as the application performs tests on increasing matrix sizes. On the other hand, the effect of our fault programs, marked in gray, can be clearly observed in the system: the \emph{cpufreq} fault causes a sudden drop in CPU frequency, resulting in reduced performance and longer computation times, while the \emph{leak} fault causes a steady, linear increase in RAM usage. Even though saturation of the available RAM is not reached, this peculiar behavior of the leak fault can be used for prediction purposes, opening up the possibility of preventing failure events.

\subsection{Overhead of FINJ}
\label{section:overhead}

We also performed tests in order to evaluate the overhead that FINJ may introduce. To do so, we employed the same system used in Section \ref{section:execution} together with the HPL benchmark, this time configured to use all 16 cores of the machine. We run the HPL benchmark 20 times directly, and then repeated the same process by using a FINJ workload. FINJ was once again instantiated locally. In both conditions the HPL benchmark scored an average running time of roughly 320 seconds, therefore leading us to conclude that the impact of FINJ on running applications is negligible. This is as expected from the implementation, since FINJ is designed to perform only the bare minimum amount of operations in order to start and manage tasks, without affecting their execution.

\section{Conclusions}
\label{section:conclusions}

We have presented FINJ, a high-level, easy-to-use tool for fault injection and monitoring in HPC systems. FINJ allows for the automation of complex experiments, and for reproducing anomalous behaviors in a deterministic, simple way. FINJ is open-source and implemented in Python, an object-oriented interpreted programming language available on all major operating systems, and has no dependencies for its core operation. This, together with the simplicity of its command-line interface, makes the deployment of FINJ on large-scale systems trivial. Since FINJ is based on the use of tasks, which are external executable programs, users can integrate the tool with any existing lower-level fault injection framework that can be triggered in such way, and ranging from the application level to the kernel, or even hardware level. The use of workloads in FINJ also allows to reproduce complex, specific fault conditions in HPC systems, and to reliably perform experiments involving multiple nodes at the same time.

As future work, we plan to perform scalability studies on the FINJ tool, by deploying it on a large-scale HPC environment. We have already performed extensive testing on the system presented in Section \ref{section:casestudy} with excellent preliminary results. Also, we plan to implement the ability to build workloads in which the order of tasks is defined by \emph{causal} relationships rather than time-stamps, which might simplify the triggering of extremely specific anomalous states in a given system. We will also integrate multiple network transport methods to choose from besides TCP, so as to extend the range of systems FINJ can be applied to.

\subsubsection{Acknowledgements.} A. Netti has been supported by a research fellowship from the \textit{Oprecomp-Open Transprecision Computing} project. A. S\^irbu has been partially funded by the EU project \textit{SoBigData Research Infrastructure --- Big Data and Social Mining Ecosystem} (grant agreement 654024).

\bibliographystyle{splncs03}
\bibliography{main}

\end{document}